\documentclass[prl,twocolumn,showpacs,preprintnumbers,amsmath,amssymb,floatfixX]{revtex4}

\usepackage{graphicx}
\usepackage{epstopdf}
\usepackage{dcolumn}
\usepackage{bm}

\begin{document}

\title{
Possibility of ``magic'' trapping of three-level system for Rydberg blockade implementation
}
\author{Muir J. Morrison}
\affiliation{
Department of Physics, University of Nevada, Reno, Nevada 89557}

\author{Andrei Derevianko}
\affiliation {
Department of Physics, University of Nevada, Reno, Nevada 89557}

\date{\today}

\begin{abstract}
The Rydberg blockade mechanism has shown  noteworthy promise for scalable quantum computation with neutral atoms.
Both qubit states and gate-mediating Rydberg state belong to
the same optically-trapped atom. The trapping fields, while being essential, induce detrimental decoherence. Here we theoretically demonstrate that this Stark-induced decoherence may be completely removed using powerful concepts  of ``magic'' optical traps.  We analyze ``magic'' trapping of a prototype three-level system: a Rydberg state along with two qubit states: hyperfine states attached to a $J=1/2$ ground state. Our numerical results show that, while such a ``magic'' trap for alkali metals would require prohibitively large magnetic fields, the group IIIB metals such as Al are suitable candidates.
\end{abstract}


\maketitle


Multiparticle quantum gates have been successfully implemented in neutral atoms using the Rydberg blockade mechanism~\cite{JakCirZol00,LukFleCot01,SafWalMol10,UrbJohHen09}. In this method, qubits are encoded in hyperfine sublevels of the electronic ground state in each atom. Gate operations utilize interactions between pairs of neutral atoms with a separation on the order of 10~$\mu$m. At this range, ground state atoms have negligible interactions, while Rydberg states interact strongly with each other. Blockade refers to the inhibition of excitation of a target atom by the prior excitation of another (control) atom. This can be used to entangle the qubits and execute CNOT gates, which in principle forms a universal set of two-qubit gates for arbitrary quantum computation.

Unfortunately, such a qubit implementation suffers from decoherence due to the trapping lasers. All three (two qubit and Rydberg) atomic levels are shifted via the Stark effect, leading to undesired differential phase accumulation. Even worse, the shift accumulation is uncontrollable as it depends on the local laser intensity, which varies as the atom moves in the trap.

\begin{figure}
\begin{center}
\includegraphics*[scale=0.28]{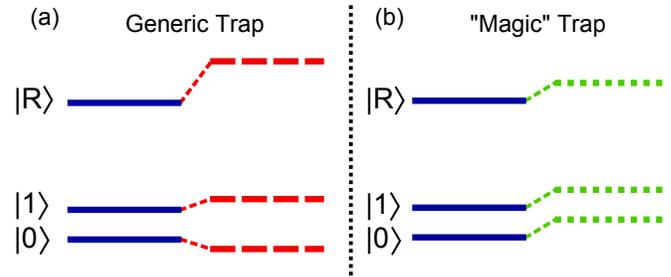}
\end{center}
\caption{(Color online)
Stark shifts affecting the Rydberg blockade three-level system. Qubit levels $|0\rangle$ and $|1\rangle$ are hyperfine levels attached to the electronic ground state, while $|R\rangle$ is a Rydberg state. The unperturbed levels (solid) acquire a shift from the trapping laser. 
(a) For an arbitrary trapping laser, each state is shifted (dashed) by a different amount according to the local laser intensity. (b) With ``magic" trapping conditions, all levels experience the same 
shift (dotted), and differential phase accumulation due to the trap vanishes.
 \label{Fig:Stark cartoon}
 }
\end{figure}

This problem is similar to that encountered in optical lattice clock experiments. So-called ``magic'' traps were proposed as a powerful solution and have seen widespread use in clock experiments (see, e.g., reviews~\cite{DerKat11,YeKimKat08} and also~\cite{KatTakPal03,RosGheDzu09}). Just as for neutral atom implementation of qubits, clocks depend on the transition frequency between two particular atomic energy levels. With a particular choice of trapping laser wavelength (and, in some cases, magnetic fields and polarization, see~\cite{DerKat11,MorDzuDer11}), both levels experience the same shift regardless of laser intensity, so the perturbation from the trap effectively vanishes. Fig.~\ref{Fig:Stark cartoon} depicts the level structure.

In the case of the alkalis, qubit transitions can be made Stark- and Zeeman-insensitive at a range of trapping wavelengths using a bias magnetic field tuned to a ``magic" value, as described in~\cite{Der10DoublyMagic}.
The hyperfine structure of Al is also favorable for ``magic" trapping, discussed in~\cite{MorDzuDer11} and~\cite{BelDerDzu08Clock}, and the formalism from~\cite{Der10DoublyMagic} for alkalis is also applicable to Al. Thus, the problem of removing decoherence for the qubit levels has already been solved.

However, excitation to a Rydberg level also causes decoherence of the qubit since in general the Rydberg trapping potential differs from the ground state potential. This third level and additional source of decoherence must be addressed. Previous proposals have considered ``magic" trapping for a two-level system including a Rydberg state in alkalis~\cite{SafWilCla03} and in alkaline-earths~\cite{OvsDerGib11,MukMilNat11}. However, such treatments are not sufficient for the present problem, as the remaining Stark shifts of the qubit states would cause decoherence. We must therefore treat all three levels simultaneously. 
This problem is more akin to the situation in~\cite{MorDzuDer11}, which considered two-level systems in two species sharing the same trap.

Our goal in this paper is to find ``magic" trapping conditions such that all three levels experience the same optical trapping potential. We begin by reviewing the Stark effect theory underlying such ``magic'' traps. We then present results of numerical calculations for Rb, Cs, and Al.

\emph{Non-magnetic states} ---
We begin by summarizing the Stark effect formalism necessary for Rydberg states. This treatment is similar to that in~\cite{ManOvsRap86,MorDzuDer11} and references therein. 
The ac Stark shift of an atomic state $| nF, M_F\rangle$ with total angular momentum $\vec{F}=\vec{J}+\vec{I}$ and projection $M_F$ can be written as~\footnote{We use atomic units throughout unless noted otherwise.}
\begin{equation}
\delta E_{nFM_F}(\omega)= - \alpha_{nFM_F}^{tot}(\omega) \left(\frac{\mathcal{E}_L}{2}\right)^{2} \, ,
\label{Eq:dE}
\end{equation}
where $\mathcal{E}_L$ and $\omega$ are the amplitude and frequency of the trapping laser field. The pre-factor $\alpha_{nFM_F}^{tot}(\omega)$ is the total polarizability of the state; note it depends on $\omega$ but not on $\mathcal{E}_L$. $\alpha_{nFM_F}^{tot}(\omega)$ may be decomposed using irreducible tensor operators as
\begin{widetext}
  \begin{equation}
      \alpha_{nFM_F}^{tot}(\omega)= \alpha_{nF}^S (\omega)
    + (\hat{k}\cdot\hat{B}) \, \mathcal{A} \, \frac{M_F}{2F}\alpha_{nF}^a (\omega)
       + \frac{1}{2}\left(3\left|\hat{\varepsilon}\cdot\hat{B}\right|^{2}-1\right)
        \frac{3{M_F}^{2}-F(F+1)}{F(2F-1)}\alpha_{nF}^T (\omega) \label{Eq:alphaTot} \, .
  \end{equation}
\end{widetext}
Here $\mathcal{A}$ is the degree of circular polarization $(|\mathcal{A}|\le1)$, while $\alpha_{nF}^S$, $\alpha_{nF}^a$, and $\alpha_{nF}^T$ are the irreducible scalar, vector, and tensor polarizabilities, respectively. The unit vectors are the laser wavevector ($\hat{k}$), laser polarization ($\hat{\varepsilon}$), and bias magnetic field ($\hat{B}$). Their relative geometry is as in~\cite{MorDzuDer11}.
%
%
The bias magnetic field is a static, externally applied field which defines the quantization axis. 
This ``quantizing magnetic field" guarantees that $M_F$ remains a ``good" quantum number for the ac Stark effect perturbation formalism, from which Eq.~(\ref{Eq:alphaTot}) follows~\footnote{The energy shifts caused by the Stark effect must be small compared to the Zeeman splitting of the magnetic sublevels.}. For linearly polarized light, 
$\hat{\varepsilon}\cdot\hat{B} = \cos{\theta_p}$, where $\theta_p$ is the angle between the polarization and quantization unit vectors. If we consider circularly polarized light, defining $\hat{\varepsilon}$ using Jones calculus conventions, then 
$\left|\hat{\varepsilon}\cdot\hat{e}_z\right|^{2} = \frac{1}{2}\sin^{2}{\theta_k}$, where $\theta_k$ is the angle between the wavevector and quantization unit vectors.
%
%

Eqs.~(\ref{Eq:dE}) and~(\ref{Eq:alphaTot}) determine our task: we must find $\omega$ and either $\theta_k$ or $\theta_p$ such that the Rydberg polarizability $\alpha_{Ryd} (\omega)$ and the qubit state polarizabilities $\alpha_{nFM_F} (\omega)$ and $\alpha_{{nF^\prime}M_{F^\prime}} (\omega)$ are all equal. Note that the polarizabilities in Eqs.~(\ref{Eq:dE}) and~(\ref{Eq:alphaTot}) may be the conventional second-order quantities, as in~\cite{ManOvsRap86}, or they may be replaced with third-order hyperfine mediated polarizabilities, denoted $\beta$ and described in detail in~\cite{RosGheDzu09}. The second-order definition considers only the interaction with the external electric field, neglecting the hyperfine interaction. This is sufficient for matching Rydberg and ground state polarizabilities, but it cannot be used to find ``magic" conditions for the qubit levels. As shown in Ref.~\cite{RosGheDzu09}, if the hyperfine interaction is neglected, the qubit levels are degenerate and always experience the same shift, so any choice of $\omega$ and $\theta$ is trivially ``magic." Therefore, a third-order treatment including hyperfine and external electric field interactions is necessary for the qubit states.

For Rydberg states, the 
tensor polarizability is highly suppressed at optical frequencies~\cite{ManOvsRap86,SafWilCla03}, so the total polarizability is dominated by the scalar part, which is essentially equal to that of a free electron, $\alpha^S_{Ryd} (\omega) = -1/\omega^2$. Our \textit{ab initio} numerical calculations confirmed this conclusion; see 
below for details. Since $\alpha^S_{Ryd}$ has negligible dependence on geometry (either $\theta_k$ or $\theta_p$), the ``magic" trap frequency will be determined by $\alpha^S_{Ryd}(\omega) = \alpha^{tot}_{nl_{1/2}}(\omega)$, i.e., when the ground and Rydberg state polarizabilities are equal.

Eqs.~(\ref{Eq:dE}) and~(\ref{Eq:alphaTot}) also apply for qubit states in Al, with $\beta$ in place of $\alpha$ as detailed above. Using Al in a linearly polarized trap, ``magic" conditions for qubit transitions between nonmagnetic states will be set entirely by geometry. 
For the alkalis, ``magic" trapping is not possible using nonmagnetic states, so we consider magnetic states below.

\emph{Magnetic states} ---
Here we review the formalism to find ``magic" conditions for qubit states with nonzero $M_F$ projection. 
The methods developed above 
are not sufficient to find ``magic" conditions for the qubit transition in alkalis. As was shown in~\cite{RosGheDzu09}, ``magic" conditions do not exist for the nonmagnetic hyperfine states. This is due to the smallness of $\alpha^T$ in comparison with $\alpha^S$, along with the strict proportionality of $\alpha^S_{{nF^\prime}M_{F^\prime}}$ and $\alpha^S_{nFM_F}$. However, if we move to magnetic substates (with $M_F\ne0$) we acquire unacceptable Zeeman sensitivity.

A solution to this problem was developed in~\cite{Der10DoublyMagic}. Atoms with a $J=1/2$ ground state are held in a circularly polarized trap to take advantage of vector polarizabilities. By utilizing multiphoton transitions between magnetic states with opposite projections (i.e., $|nF^\prime,M_F\rangle$ and $|nF, -M_F\rangle$) most of the first order Zeeman shift goes away as these states have opposite electronic g-factors \footnote{Since $J=1/2$, there are only 2 hyperfine states, and $F^\prime=F+1.$}. The remaining first-order shift is due only to the much smaller nuclear magnetic moment, which can be made to cancel the second-order shift with the application of a static magnetic field. The ``magic" value of the B-field is given by
\begin{equation}
 B_{m}\approx
\frac{g_{I}\mu _{N}~M_{F^{\prime }}}{2\left\vert \langle nF,M_{F^{\prime }}\left\vert \mu _{z}^{e}\right\vert nF^{\prime },M_{F^{\prime }}\rangle \right\vert ^{2}} \, \omega_\mathrm{qubit} \,,
\label{Eq:Bmagic}
\end{equation}
where $\omega_\mathrm{qubit}$ is the energy splitting between the hyperfine levels. Although this expression for $B_{m}$ is only accurate to second-order, it agrees well with the exact answer~\cite{Der10DoublyMagic}.

Having dealt with Zeeman sensitivity, we turn to the Stark shift due to the trapping lasers. We need the differential shift between the qubit states to vanish, i.e., both states must experience the same shift. This condition is satisfied when~\cite{Der10DoublyMagic}
\begin{widetext}
\begin{equation}
\left(  \beta_{nF^{\prime}}^{s}-\beta_{nF}^{s}\right)  + \delta\beta^T +
  \mathcal{A}\cos\theta_k~M_{F^\prime} \left[  \left(  \frac
{1}{2F^{\prime}}\beta_{nF^{\prime}}^{a}+\frac{1}{2F}\beta_{nF}^{a}\right)
+
g_{I}\frac{\mu_{N}}{\mu_{B}}\left(\frac{B}{B_{m}}\right)~{\overline{\alpha}}^{a}_{nl_{1/2}}
 \right] = 0  
 . \label{Eq:ClockShiftMultiPhoton}
\end{equation}
\end{widetext}

Here $\beta^S$, $\beta^a$, and $\beta^T$ are the third-order hyperfine mediated polarizabilities referred to above, while $\overline{\alpha}^a_{nl_{1/2}}$ is the conventional second-order polarizability. Note the distinction between coupling schemes used in $\overline{\alpha}^a_{nl_{1/2}}$ and $\alpha_{nF}^a$. $\delta\beta^T$ absorbs lengthy prefactors. The full form is
\begin{equation}
\delta\beta^T = -\beta^T_{nF^\prime}\frac{3M_F^2-F^\prime(F^\prime+1)}{2F^\prime(2F^\prime-1)} + \beta^T_{nF}\frac{3M_F^2-F(F+1)}{2F(2F-1)}.
\end{equation}
$B_m$ is set by Eq.~(\ref{Eq:Bmagic}), while $B$ is the actual applied field. Ideally $B=B_m$, but as suggested in~\cite{Der10DoublyMagic} this is not always possible. 

Clearly we need the third-order polarizabilities for the two states to cancel, but the appearance of the last term involving $\alpha^a_{nS_{1/2}}$ may be surprising. It arises due to interference between the Zeeman shift and the vector part of the Stark shift, as these are both (axial) vector operators; see~\cite{Der10DoublyMagic} for a full discussion. Although $\mu_N \ll \mu_B$, $\alpha \gg \beta$, so this term is of a comparable order of magnitude and must be included in Eq.~(\ref{Eq:ClockShiftMultiPhoton}).


\emph{Numerical evaluation} ---
We used the same codes as in~\cite{MorDzuDer11} for the ground states in all atoms considered. To summarize, we use the B-spline technique to generate a quasi-complete set of orbitals that are solutions to the Dirac-Hartree-Fock equations. To refine these solutions, we find the second-order self-energy operator to build the so-called Brueckner orbitals. Matrix elements are then calculated using the relativistic random-phase approximation.

For the Rydberg state calculations, we extended and modified our B-spline codes; the original codes are described in detail in~\cite{BelDer08}. To generate a complete basis set including physically accurate Rydberg states, we dramatically increased the size of the cavity and the number of basis functions. As an illustration, a typical run to calculate low-lying states uses $\sim40$ splines in a $\sim50a_B$ cavity. For calculations aimed at the $50s$ state, we obtained accurate results using $\sim200$ splines in a $\sim8000a_B$ cavity. We also used a logarithmic rather than an exponential distribution of spline knots. This increased the accuracy of matrix elements by improving the representation of wavefunctions at large $R$ near the cavity wall.

Correlations were included by building the self-energy operator using a small basis set ($\sim40$ splines in a $\sim50a_B$ cavity) and using this potential to build Brueckner orbitals for a large set. This is justified since the self-energy operator diminishes rapidly outside the core, so highly excited states have a negligible contribution. Neglecting these states decreases calculation time dramatically. Inclusion of correlations introduced small but detectable corrections to Rydberg state energies and matrix elements. Corrections were around the fourth significant figure for $n=50$ states and diminished with increasing $n$.

\begin{figure} [h]
\begin{center}
\includegraphics*[scale=0.82]{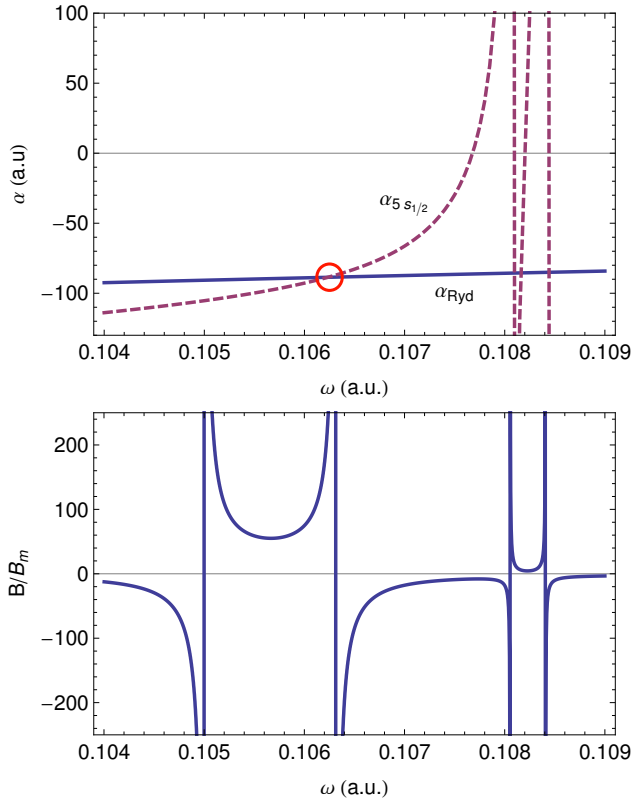}
\end{center}
\caption{(Color online)
Polarizabilities of the 5s state (dashed), the Rydberg state (solid), and the ratio $B/B_m$ (lower frame) for the hyperfine transition $| F=2,M_F=1\rangle$ to $| F=1,M_F=-1\rangle$ in $^{87}$Rb. Since $\alpha^a_{5s} \ll \alpha^S_{5s}$, the ``magic" $\omega$ simply occurs where $\alpha_{Ryd} = \alpha^S_{5s_{1/2}}$ at approximately $\omega = 0.1062$~a.u. ($\lambda=429$~nm). $B/B_m$ is obtained from Eq.~(\ref{Eq:ClockShiftMultiPhoton}). Near the circled ``magic" $\omega$, $B/B_m$ diverges, so magic trapping is impossible.
 \label{Fig:Rb_Bmagic}
 }
\end{figure}

\emph{Results} ---
We first present results for $^{87}$Rb. This isotope has nuclear spin $I=3/2$, and we are interested in transitions between the $| F^{\prime}=2, M_{F^\prime}=1\rangle$ and  $| F=1, M_F=-1\rangle$ qubit states attached to the $5s_{1/2}$ electronic ground state. The hyperfine splitting is $6.83$~GHz, and from Eq.~(\ref{Eq:Bmagic}), $B_m \approx 3.25$~G~\cite{Der10DoublyMagic}. In Fig.~\ref{Fig:Rb_Bmagic} we plot $\alpha_{Ryd}$, $\alpha^S_{5s_{1/2}}$, and the ratio $B/B_m$. To achieve ``magic" trapping using the scheme above, 
we require $| B/B_m | \le 1$ in Eq.~(\ref{Eq:ClockShiftMultiPhoton}). If $| B/B_m |$ is slightly greater than 1, ``nearly-magic" trapping is possible as considered in~\cite{Der10DoublyMagic}. But in this case, $| B/B_m |$ diverges near $\omega_m$, and 
the necessary $B$ to achieve Stark-insensitive trapping is prohibitively large. 
The situation is qualitatively the same for $^{133}$Cs; our numerical calculations show a similar divergence near $\omega_m$. For the lighter alkalis, there are no doubly-magic or nearly-magic points for the qubit transition.

Since it is not possible to build a ``magic" three-level trap with the alkalis, we turn to $^{27}$Al. As was shown in~\cite{BelDerDzu08Clock}, ``magic" trapping of its hyperfine states is aided by comparatively large vector and tensor polarizabilities. With nuclear spin $I=5/2$, we consider the $| F^{\prime}=3, M_{F\prime}=1\rangle$ and  $| F=2, M_F=-1\rangle$ qubit states attached to the $3p_{1/2}$ electronic ground state. The hyperfine splitting is $1.506$~GHz, and Eq.~(\ref{Eq:Bmagic}) gives $B_m \approx 4.32$~G. We may use either a linearly or circularly polarized trapping laser. In the case of a linearly polarized trap, the situation is closely analogous to that in~\cite{MorDzuDer11}. The ``magic" wavelength is set where $\alpha_{Ryd} = \alpha^S_{3p_{1/2}}$, near $\omega = 0.121$~a.u. ($\lambda = 377$~nm). The enhanced tensor polarizability all but guarantees that the qubit transition can be made ``magic." Numerical calculations show this does occur with a ``magic" angle $\theta_p \approx 65^{\circ}$.

\begin{figure} [h]
\begin{center}
\includegraphics*[scale=0.8]{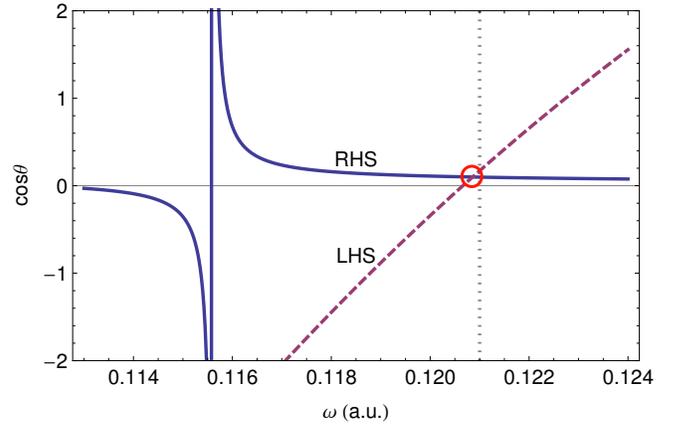}
\end{center}
\caption{(Color online)
Left-hand side (dashed) and right-hand side (solid) of Eq.~(\ref{Eq:AlCirc}). The range for ``magic" trapping of the Rydberg transition is given by Eq.~(\ref{Eq:alphaAl}) and lies between the resonance at $\omega = 0.1155$ a.u. ($\lambda = 394$~nm) and the dotted line at $\omega = 0.121$ a.u. ($\lambda = 377$~nm). The ``magic" frequency and ``magic" angle for the qubit transition are set by the curves intersection, just under $\omega = 0.121$ a.u. This combination of $\omega_{magic}$ and $\theta_{magic}$ will allow Stark and Zeeman insensitive trapping for the three-level system in Al.
 \label{Fig:AlThetaMagic}
 }
\end{figure}

For a circularly polarized trap, 
no divergence of $B/B_m$ occurs near the ``magic" frequency $\omega_m$ as for the alkalis. However, the ``magic" condition is more complex, as the ground state second order $\alpha^a_{3p_{1/2}}$ is not negligible compared with $\alpha^S_{3p_{1/2}}$. The total second-order polarizability of the ground state is given by Eq.~(\ref{Eq:alphaTot})~\footnote{$\alpha^T$ is identically zero as $J=1/2$.}. This must equal the Rydberg polarizability, leading to
  \begin{equation}
       \alpha_F^S (\omega)
    + \mathcal{A} \, \cos \theta_k \frac{M_F}{2F}\alpha_F^a (\omega) =  \alpha_{FM_F}^{Ryd}(\omega).
\label{Eq:alphaAl}
  \end{equation}
Our choice of $\omega$ and $\theta_k$ must simultaneously satisfy Eqs.~(\ref{Eq:ClockShiftMultiPhoton}) and (\ref{Eq:alphaAl}). We may solve for $M_F\mathcal{A}\cos\theta_k$ appearing in Eqs.~(\ref{Eq:ClockShiftMultiPhoton}) and~(\ref{Eq:alphaAl}) and equate the results, giving
\begin{equation}
     2F^{\prime}\frac{\omega^{-2}+\alpha_F^S}{\alpha_F^a}
=  \frac {
              \beta_{F^{\prime}}^{s}-\beta_{F}^{s} + \delta\beta^T
            }
           {
             \frac{1}{2F^{\prime}}\beta_{F^{\prime}}^{a}+\frac{1}{2F}\beta_{F}^{a} +
                  g_I\frac{\mu_N}{\mu_B}\frac{B}{B_m}\overline{\alpha}_{np_{1/2}}^{a}
	    }.
\label{Eq:AlCirc}
\end{equation}

We plot the left and right hand sides of Eq.~(\ref{Eq:AlCirc}) in Fig.~\ref{Fig:AlThetaMagic}. Since they intersect in the range allowed by Eq.~(\ref{Eq:alphaAl}), ``magic" trapping for this three-level system in Al is possible. While circularly polarized trapping would be more complex than the linearly polarized trap presented above, it has the additional advantage of less Zeeman sensitivity. This is because the ``magic" $B$-field removes Zeeman effects to second-order, while the linearly polarized trap only removes Zeeman decoherence to first order in the $B$-field.

We have presented a method to remove differential Stark shifts for a three-level atomic system consisting of a Rydberg state and two hyperfine states attached to the ground electronic state. Such system is an essential element of the CNOT gates utilizing Rydberg blockade mechanism.
Although such a trap is not possible for the alkalis, our numerical calculations show that Al may be trapped using this method. Such a trap could prove useful for removing decoherence from trapping lasers in implementing the Rydberg blockade mechanism.

\emph{Acknowledgements ---}
  We would like to thank J.\ Weinstein and M.\ Saffman  for  discussions.  This work was supported in part by the NSF.


\end{document}